\documentstyle[prc,aps,psfig]{revtex}
\begin{document}
\twocolumn

\title{\Large\bf On the discovery of doubly-magic $^{48}$Ni}
 
\author{ 
B. Blank, M. Chartier, S. Czajkowski, J. Giovinazzo, M.S.~Pravikoff, J.-C. Thomas} 

\address{
CEN~Bordeaux-Gradignan, Le Haut-Vigneau, F-33175 Gradignan Cedex, France}

\author{ 
G. de France, F. de Oliveira Santos, M. Lewitowicz }

\address{
Grand Acc\'el\'erateur National d'Ions Lourds, B.P. 5027, F-14076 Caen Cedex, 
France}

\author{ 
C. Borcea}

\address{
IAP, Bucharest-Magurele, P.O. Box MG6, Rumania}

\author{ 
R. Grzywacz{\footnotemark}, Z. Janas, M. Pf\"utzner }

\address{
Institute of Experimental Physics, University of Warsaw, PL-00-681 Warsaw, 
Hoza 69, Poland }

\maketitle

\begin{abstract}
In an experiment at the SISSI/LISE3 facility of GANIL, we used the projectile 
fragmentation of a primary $^{58}$Ni$^{26+}$ beam at 74.5~MeV/nucleon with an 
average current of 3~$\mu$A on a natural nickel target to produce very 
neutron-deficient isotopes. In a 10-day experiment, 287 $^{42}$Cr
isotopes, 53 $^{45}$Fe isotopes, 106 $^{49}$Ni isotopes, and 4 $^{48}$Ni 
isotopes were unambiguously identified. The doubly-magic nucleus $^{48}$Ni, 
observed for the first time in the present experiment, is the most proton-rich 
isotope ever identified with an isospin projection $T_z$~= -4. At the 
same time, it is probably the last doubly-magic nucleus with "classical" 
shell closures accessible for present-day facilities. 
Its observation allows us to deduce a lower limit for the half-life of 
$^{48}$Ni of 0.5~$\mu$s.

\end{abstract}

\vspace*{0.2cm}
\vspace*{0.2cm}

\renewcommand{\thefootnote}{\fnsymbol{footnote}}
\footnotetext[1]{present address: Department of Physics and Astronomy, 
University of Tennessee, Knoxville, TN 37996-1200, USA}

Doubly magic nuclei have a magic attraction for nuclear physicists. These nuclei
enabled them to interpret many properties of nuclei. They are spherical
and exhibit a remarkable stability compared to neighboring isotopes. However, 
one of the questions of current interest is whether or not the role of the magic
numbers, well established along the valley of stability, 
remains important when an extreme excess of protons or neutrons is 
present in a nucleus. In fact, lighter naturally occuring isotopes have roughly 
the same number of protons and neutrons. Thus the lightest stable isotope of the 
element nickel is $^{58}$Ni with 28 protons and 30 neutrons, while $^{57}$Ni is 
already unstable and decays by $\beta^+$ emission. The excess of protons or 
neutrons in a nucleus is characterized by the isospin-projection quantum number 
$T_z$ which is half the difference between neutron and proton numbers. The most 
negative value for $T_z$ is expected to occur for $^{48}$Ni.

The recent observation of doubly-magic
$^{100}$Sn~\cite{schneider94,lewitowicz94} and $^{78}$Ni~\cite{engelmann95} 
showed that projectile fragmentation is most probably the route to $^{48}$Ni 
and the question about its stability
attracted the attention of experimentalists as well as of
theorists~\cite{brown91,ormand96,ormand97,cole96,nazarewicz96}. 
The special interest of this nucleus is that it is the only case
of a doubly-magic nucleus in the entire chart of nuclei where the mirror 
nucleus, $^{48}$Ca, is stable 
against particle emission, a fact which may allow for interesting mirror 
symmetry studies. In addition, with $^{48}$Ni, $^{56}$Ni, and $^{78}$Ni, 
nickel is most probably the only element possessing three doubly-magic nuclei. 
Furthermore, together with $^{45}$Fe, $^{48}$Ni is the prime candidate for 
the yet unobserved two-proton radioactivity, the emission of $^2$He 
from the ground state of an isotope~\cite{brown91,ormand96,ormand97,cole96,nazarewicz96}.
If one excludes the super-heavy elements, where the magic numbers are 
not known experimentally, $^{48}$Ni is also predicted to be the last 
doubly-magic nucleus possessing shell closures known from spherical nuclei
which is accessible with present-day techniques.

The search for doubly-magic $^{48}$Ni started with the advent of powerful
projectile-fragmentation facilities. Thus, $^{51,52}$Ni have been identified 
for the first time in 1986~\cite{pougheon87} in an experiment at the LISE 
facility~\cite{lise}. In this experiment, an average beam current of 50~nA 
corresponding to 1.5$\times$10$^{10}$ particles per second for the 
primary $^{58}$Ni beam at 55~MeV/nucleon was available. 

The next two steps towards $^{48}$Ni were taken at the fragment separator 
FRS~\cite{frs} at GSI. In a first experiment using the projectile fragmentation 
of a relativistic $^{58}$Ni beam at 650~MeV/nucleon (5$\times$10$^7$ particles 
per second) in 1993~\cite{blank94ni}, 
$^{50}$Ni was observed with 3 events. At the same time, extensive production
cross-section measurements were performed which enabled us for the first time
to reliably predict the production cross section of proton-rich isotopes from 
$^{58}$Ni fragmentation (see below). After an increase of the primary-beam 
intensity of about a factor of 10 from SIS at GSI, a second 
experiment~\cite{blank96fe45} was carried out at the FRS using the projectile 
fragmentation of a primary $^{58}$Ni beam at 600~MeV/nucleon to produce
proton-rich isotopes. The experiment led to the discovery of $^{42}$Cr, 
$^{45}$Fe, and $^{49}$Ni.

Intensive developments on the ECR ion sources at GANIL made it possible to start
a search for doubly-magic $^{48}$Ni. In these developments, the metalic 
natural nickel used in the past was replaced by nickelocene, a chemical 
compound which allowed to treat nickel as a gas in the ECR ion source.

A primary beam of $^{58}$Ni$^{26+}$ with an average current of 3~$\mu$A corresponding 
to 7.2$\times$10$^{11}$ particles per second was accelerated by the GANIL cyclotrons 
to an energy of 74.5~MeV/nucleon. This beam impinged on a natural nickel target
of thickness 230.6~mg/cm$^2$ followed by a 2.7~mg/cm$^2$ carbon stripper foil,
both installed on a rotating wheel between the two superconducting solenoids of
the SISSI device. The secondary beams (B$\rho_1$~= 1.7323~Tm) were transported 
through the Alpha spectrometer to the LISE3 separator~\cite{lise}. At the 
intermediate focal plane of LISE3, a shaped beryllium degrader (10.36~mg/cm$^2$)
in conjuntion with the second dipole stage (B$\rho_2$~= 1.7000~Tm) allowed for 
a refined selection of the fragments of interest. 

At the first achromatic focal plane of LISE3, a micro-channel-plate (MCP) detector 
was mounted to yield a timing signal for each isotope. This detector had to 
withstand a counting rate of about 5$\times$10$^5$ particles per second. 
It turned out that its efficiency was about 70\%.
The velocity filter at the end
of the LISE3 device reduced the counting rate to roughly 500 particles per
second. The selected isotopes were finally implanted in a silicon-detector telescope
consisting of five detectors, the first three with a thickness of 300~$\mu$m each, 
the fourth with a thickness of 700~$\mu$m and the last one with a thickness of 6~mm. 
The first detector (active area 600~mm$^2$) was used for an energy-loss and a 
time-of-flight measurement, whereas the second detector (quadratic shape with an 
area of 3$\times$3~cm$^2$) was position-sensitive (resistive read-out) and thus 
yielded an energy-loss and a two-dimensional position measurement with a 
spatial resolution of about 1~mm.
The isotopes of interest were finally implanted in the third detector (600~mm$^2$)
which measured the residual energy and yielded another timing signal.
Most of the lighter or less exotic particles were stopped 
in the first two silicon detectors. 
Thus the counting rate in the implantation detector was only about 10-15 
particles per second. The last two detectors served as veto detectors and 
had a size of 15$\times$40~mm$^2$ and of 600~mm$^2$. 

This detection set-up enabled us to measure 10 independent
parameters to identify the isotopes for each implantation event: 
three energy signals in the first three silicon detectors, two veto 
signals in the fourth and the fifth detector, positions in x 
and y, and three independent flight times (TOF). The TOFs were 
measured between the MCP detector and the first silicon detector, the first silicon
detector and the cyclotron radiofrequency, as well as between the third silicon 
detector and the radiofrequency.

To analyse the data, the average value for each of these 10 parameters as well as 
its full width at half maximum (FWHM) was determined 
for each isotope based on the experimental observation.
In order to calculate the parameters for particle-unstable isotopes 
not present in the spectra, interpolations and
extrapolations from the measured values of neighboring nuclei were used.
The expected average value for isotopes with isospin projections from
$T_z$~= -2 to $T_z$~= -9/2 and with nuclear charges from $Z$~= 22 to $Z$~= 29 was 
calculated for each parameter. To accept an event, all parameters had to 
lie within a window of two FWHM. However, for the graphical representation, these
cuts were only applied for parameters not represented in the spectrum. 
Figure~1 shows the result
of this analysis as a plot of the energy loss in the first silicon detector
as a function of the TOF between the MCP detector and the first silicon detector.
Due to the efficiency of the MCP detector, part of the statistics is lost in this 
spectrum. Nevertheless, two events of $^{48}$Ni are clearly observed. In 
particular, no background events are observed in the whole region analysed between the
unbound $^{54}$Cu at the upper left end and the unbound $^{35}$Ti
at the lower right end. In addition, 77 events of 
$^{49}$Ni, 29 events of $^{45}$Fe, and 164 events of $^{42}$Cr fullfil the stringent
conditions imposed on this spectrum. Already from this spectrum we conclude 
therefore that the new doubly-magic nucleus $^{48}$Ni has been unambiguously
identified.

\begin{figure}[h]
\psfig{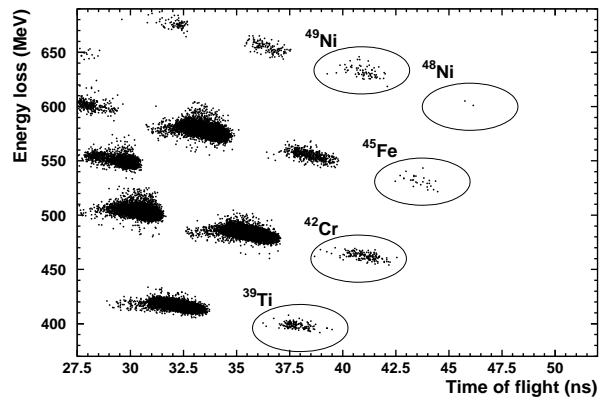}
\caption{Two-dimensional identification plot of energy loss in the first silicon 
         detector versus the TOF between the MCP detector and the silicon detector. 
         To produce this spectrum, software conditions were applied to all
         parameters except those represented here (see text for details). 
         Due to the efficiency of the MCP detector, part of the events are lost
         in this spectrum.}
\protect\label{fig1}
\end{figure}

To circumvent the lack of efficiency of the MCP detector, we analysed the data
also as a function of the energy loss in the first silicon detector 
and the total energy as the sum of the signals in the first three silicon 
detectors. For this
analysis, we used the information from the MCP detector only, if it was present, 
i.e. if the TOF signal of this detector and the first silicon detector 
was above zero. The result of this analysis is shown in figure~2. In this spectrum,
we observe four counts which we can unambiguously attribute to $^{48}$Ni.
The number of counts for $^{49}$Ni, $^{45}$Fe, and $^{42}$Cr are 89, 48, and 264, 
respectively.

This spectrum, however,
demonstrates also the importance of the MCP detector, as all background counts
are events without MCP timing. In fact, we identify one background count 
at the positions of the unbound $^{51}$Cu, $^{45}$Mn, $^{42}$V, and $^{41}$V, 
whereas the other counts could be rejected due to their wrong energy-loss/energy
relation in this spectrum. However, none of these background counts are 
accumulated in a particular region in this energy-loss/energy matrix, like it is 
the case for the four $^{48}$Ni events. Therefore, it seems rather unlikely 
that more than one of these four events is background.

\begin{figure}[h]
\psfig{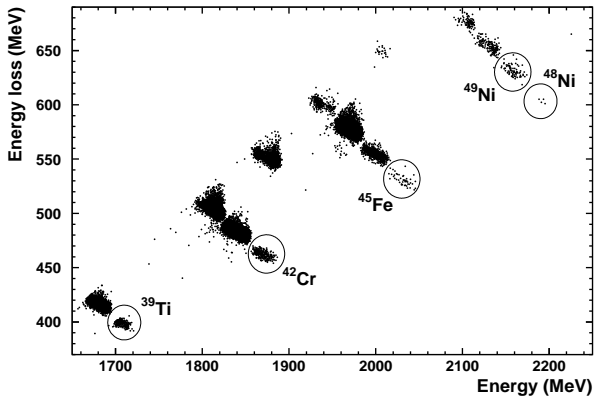}
\caption{Two-dimensional identification plot of energy loss in the first 
         silicon detector versus the total energy calculated as the sum 
         energy of the first three silicon detectors. As for figure~1, 
         software conditions were applied  to all parameters except those
         represented here (see text for details). Less exotic 
         isotopes were cut either by software conditions or by the 
         transmission of LISE3.
}
\protect\label{fig2}
\end{figure}

From figures 1 and 2, we conclude that the new doubly-magic nucleus $^{48}$Ni
has been observed for the first time. The observation of $^{49}$Ni, 
$^{45}$Fe, and $^{42}$Cr
confirms the results from the 1996 GSI experiment~\cite{blank96fe45} where
these isotopes were observed for the first time with much lower counting 
rates.

The number of counts of the present experiment for the 
four most proton-rich isotopes
$^{48}$Ni, $^{49}$Ni, $^{45}$Fe, and $^{42}$Cr is given in table~1. 
In order to have all counts for each nucleus to calculate the production 
cross sections, we used 
three-FWHM windows to determine the number of counts. To estimate
their production cross section we calculated the Alpha+LISE3 transmission for
these isotopes with three different simulation codes, the programs 
LISE~\cite{lise_code}, LIESCHEN~\cite{lieschen}, and INTENSITY~\cite{intensity}.
The INTENSITY code has no feature to treat the velocity filter of the LISE3
separator. However, it turned out in the simulations with the two other codes that
the velocity-filter transmission for the four isotopes discussed here is 100\%
and thus INTENSITY calculations yield the total transmission in these special cases.
In table~1, we give the average transmission from the three simulations. The error
is estimated from the difference of the predictions. Except for $^{42}$Cr which is
close to the acceptance limit of LISE3 at the intermediate focal plane, the 
different codes predict transmissions within a window of about 50\%.

The production cross sections for $^{48}$Ni, $^{49}$Ni, $^{45}$Fe, and $^{42}$Cr 
in the present experiment can be deduced from this information, the $^{nat}$Ni 
target thickness of 230.6~mg/cm$^2$ and a total beam dose of 4.2$\times$10$^{17}$
particles. These cross sections are presented in table~1. They are 
compared to the new EPAX formula~\cite{epax_new}, an empirical 
parametrization of projectile fragmentation cross sections. We find nice agreement
between experimentally determined cross sections and predictions from EPAX, 
especially for $^{48}$Ni and $^{49}$Ni. However, the experimental production 
cross section for $^{45}$Fe is more than a factor of two lower than expected.
From the cross-section systematics measured
for the fragmentation of a $^{58}$Ni primary beam~\cite{blank94ni}, a decrease 
of a factor of two is not expected between $^{49}$Ni and $^{45}$Fe. A possible
explanation could be that part of the $^{45}$Fe activity decays in-flight
with a half-life of roughly 0.5~$\mu$s. However, we can not exclude that
the production cross section for $^{45}$Fe might be smaller than expected
or that the transmission calculations have a larger error. Therefore, 
additional information, namely
measurements of its decay as well as a measurement similar to the present 
experiment, but with a significantly shorter flight time, are necessary to 
confirm or reject this hypothesis.

$^{48}$Ni is predicted particle unstable by all commonly used mass models. 
As it is an even-Z nucleus, these models foresee that the 
one-proton emission is energetically forbidden and thus $^{48}$Ni may decay 
by two-proton emission.  However, the model predictions for the Q value for 
two-proton emission, $Q_{2p}$, vary between less than 1.0~MeV and more 
than 3.0~MeV (see figure~3). Using barrier-penetration calculations 
this gap yields partial
half-lives for two-proton decay between 10$^{-14}$~s
and 1~s. 
\begin{table}[ht]
\begin{tabular}{|l|c|c|c|c|}
\hline \rule{0pt}{1.3em}
                  & $^{48}$Ni   &  $^{49}$Ni   &  $^{45}$Fe    &   $^{42}$Cr    \\ [0.5em] \hline \rule{0pt}{1.3em}
number of counts  &  4$\pm$2    &   106$\pm$22 &   53$\pm$9    &    287$\pm$30  \\
transmission (\%) & 9.8$\pm$1.2 & 6.9$\pm$1.6  &  6.7$\pm$0.3  &  0.7$\pm$0.6 \\
cross section (pb)&0.05$\pm$0.02& 1.8$\pm$0.6  & 0.9$\pm$0.2   & 47$\pm^{285}_{22}$\\
EPAX~\cite{epax_new} (pb) & 0.06     &  1.5         &         1.8   &     69 \\
[0.5em] \hline
\end{tabular}
\caption[]{Pertinent information concerning the production cross sections of
           $^{48}$Ni, $^{49}$Ni, $^{45}$Fe, and $^{42}$Cr. The first line gives
           the number of counts experimentally observed. The numbers given 
           correspond to 3-FWHM 
           windows for each parameter (see text). The error bars are
           the sum of statistical and systematic errors. The systematic errors
           have been determined by varying the conditions used to identify the
           isotopes. The transmissions in the second line are averages from 
           three different model calculations with LISE~\cite{lise_code}, 
           LIESCHEN~\cite{lieschen}, and INTENSITY~\cite{intensity}. 
           The experimental cross sections (third line) are determined with
           a primary beam dose of 4.2$\times$10$^{17}$ particles and a 
           $^{nat}$Ni target thickness of 230~mg/cm$^2$. In addition, a dead time
           correction of 13\% is applied. Losses due to secondary reactions in target 
           and degrader are below the one percent level and are not taken into account.
           The last line gives the predictions from the new EPAX
           parametrization~\cite{epax_new}.}
\label{table1}
\end{table}
However, a closer inspection of the mass models used in figure~3 shows 
that ``local models", i.e. models which either have been tailored to calculate 
masses in the region around $^{48}$Ni~\cite{brown91,ormand96,ormand97,cole96} 
or which use local mass relations  
like the isobaric multiplet mass 
equation~\cite{benenson75} or the Garvey-Kelson mass relation~\cite{masses88},
predict $Q_{2p}$ values  
which cluster at about 1.3~MeV. Some of the ``global models" which 
consistently predict masses for a large part of the table of 
isotopes give also masses in this Q-value range~\cite{masses88}. 
However, most of them lie above 1.5~MeV and predict therefore much 
shorther half-lives. In microscopic calculations like those of Nazarewicz and 
co-workers~\cite{nazarewicz96}, the masses depend on the Skyrme force used.

\begin{figure}[h]
\begin{center}
\psfig{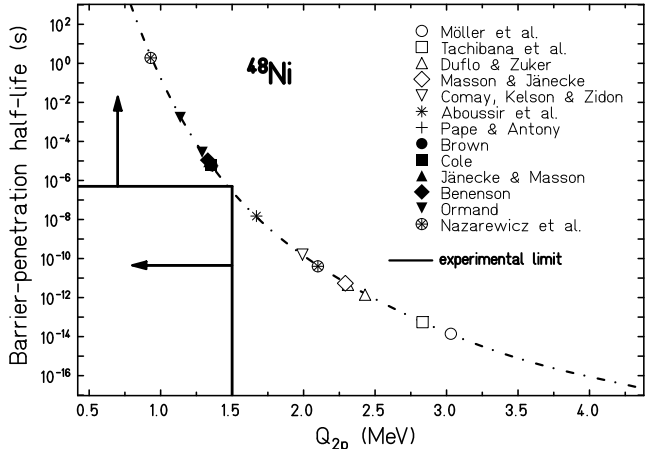}
\caption[]{Barrier-penetration half-life as a function of the two-proton Q value, 
         $Q_{2p}$, for $^{48}$Ni. The barrier penetration was calculated by 
         assuming a spectroscopic factor of unity. Different model predictions~\cite{brown91,ormand96,ormand97,cole96,nazarewicz96,benenson75,masses88,moeller95,duflo95,aboussir95}
         were used for $Q_{2p}$. The experimental observation of $^{48}$Ni 
         implies a half-life longer than 0.5~$\mu$s yielding an upper limit for 
         $Q_{2p}$ of about 1.5~MeV. 
}
\protect\label{fig3}
\end{center}
\vspace*{-0.9cm}
\end{figure}

From the experimental observation of $^{48}$Ni, we can deduce a limit of 
its half-life. Assuming that $^{49}$Ni does not decay in flight, we can use 
the number of counts observed for this isotope to estimate the number of 
counts expected for $^{48}$Ni. The transmissions for $^{48,49}$Ni
(see table~1) and a cross-section decrease of a factor 
of 26.4$\pm$4.7~\cite{blank94ni} allow 
us to deduce a number of counts expected for $^{48}$Ni of 5.7$\pm$2.2.
With these figures and a total flight time of 1.32~$\mu$s, we obtain a
half-life for $^{48}$Ni of (2.6$^{+\infty}_{-2.1})\mu$s. We determine thus a
lower limit at the 1$\sigma$ level of 0.5~$\mu$s for the life time of $^{48}$Ni.
This value translates into an upper limit for $Q_{2p}$ of about 1.5~MeV if we 
assume a spectroscopic factor of unity. Combined with the mass excess of
(0.755$\pm$0.354)~MeV for $^{46}$Fe~\cite{audi95}, we determine therefore an
upper limit for the mass excess of $^{48}$Ni of 17.2~MeV. One should keep in mind
that this determination is based on an extrapolated mass excess for $^{46}$Fe
and a spectroscopic factor of unity which make it only a rough estimate.

The $\beta$-decay half-life of $^{48}$Ni is predicted to lie within a range of 
5.9~ms and 11.3~ms~\cite{brown91,hirsch93}. Therefore, if the $Q_{2p}$ value 
is, as predicted by several mass models, of the order of 1.3~MeV, 
which yields a barrier-penetration half-life of
about 10$^{-5}$s, $^{48}$Ni should decay by two-proton ground-state
emission. Such a half-life should be accessible in future experiments.

In summary, we observed for the first time 4 events of the doubly-magic
nucleus $^{48}$Ni. The experimental observation implies that its half-life is
longer than about 0.5~$\mu$s. Using a spectroscopic factor of unity, this 
translates into an upper limit for $Q_{2p}$ of 1.5~MeV. 
Several global mass models are
in disagreement with this value, whereas local mass relations are in nice
agreement with the present observations. Future measurements of the decay
of $^{48}$Ni which should be feasible after additional improvements of the
primary-beam intensity should yield more detailed information on its mass 
and its decay properties.

We would like to acknowledge the continous effort of the GANIL accelerator 
staff to provide us with a stable, high-intensity beam. We 
express our sincere gratitude to the LISE staff for ensuring a smooth 
running of the LISE3 separator. This work was 
supported in part by the Polish Committee of Scientific Research under
grant KBN 2 P03B 036 15, the 
contract between IN2P3 and Poland, as well as by the Conseil R\'egional 
d'Aquitaine.


\vspace*{-0.35cm}
\begin{thebibliography}{10}
\vspace*{-1.25cm}

\bibitem{schneider94}
R. Schneider {\it et~al.}, Z. Phys. A {\bf 384},  241  (1994).

\bibitem{lewitowicz94}
M. Lewitowicz {\it et~al.}, Phys. Lett. B {\bf 332},  20  (1994).

\bibitem{engelmann95}
C. Engelmann {\it et~al.}, Z. Phys. A {\bf 352},  351  (1995).

\bibitem{brown91}
B.~A. Brown, Phys. Rev. C {\bf 43},  R1513  (1991).

\bibitem{ormand96}
W.~E. Ormand, Phys. Rev. C {\bf 53},  214  (1996).

\bibitem{ormand97}
W.~E. Ormand, Phys. Rev. C {\bf 55},  2407  (1997).

\bibitem{cole96}
B.~J. Cole, Phys. Rev. C {\bf 54},  1240  (1996).

\bibitem{nazarewicz96}
W. Nazarewicz {\it et~al.}, Phys. Rev. C {\bf 53},  740  (1996).

\bibitem{pougheon87}
F. Pougheon {\it et~al.}, Z. Phys. A {\bf 327},  17  (1987).

\bibitem{lise}
A.~C. Mueller and R. Anne, Nucl. Instrum. Meth. {\bf B56},  559  (1991).

\bibitem{frs}
H. Geissel {\it et~al.}, Nucl. Instrum. Meth. {\bf B70},  793  (1992).

\bibitem{blank94ni}
B. Blank {\it et~al.}, Phys. Rev. C {\bf 50},  2398  (1994).

\bibitem{blank96fe45}
B. Blank {\it et~al.}, Phys. Rev. Lett. {\bf 77},  2893  (1996).

\bibitem{lise_code}
D. Bazin, M. Lewitowicz, O. Sorlin, and O. Tarasov,   (LISE simulation code,
  unpublished).

\bibitem{lieschen}
B. Blank, E. Hanelt, and K.-H. Schmidt,   (LIESCHEN simulation code,
  unpublished).

\bibitem{intensity}
J. Winger, B. Sherrill, and D. Morrissey, Nucl. Instr. Meth. {\bf B70},  380
  (1992).

\bibitem{epax_new}
K. S{\"u}mmerer and B. Blank, Phys. Rev. C  submitted for publication  (1999).

\bibitem{benenson75}
W. Benenson, Nukleonika {\bf 20},  775  (1975).

\bibitem{masses88}
P.E. Haustein (ed.), At. Data Nucl. Data Tab. {\bf 39},  185  (1988).

\bibitem{moeller95}
P. M{\"o}ller, J.~R. Nix, W.~D. Myers, and W.~J. Swiatecki, At. Data Nucl. Data
  Tab. {\bf 59},  185  (1995).
 
\bibitem{duflo95}
J. Duflo and A. Zuker, Phys. Rev. C {\bf 52},  R23  (1995).

\bibitem{aboussir95}
Y. Aboussir, J. Pearson, A. Dutta, and F. Tondeur, At. Data Nucl. Data Tab.
  {\bf 61},  127  (1995).

\bibitem{audi95}
G. Audi and A.~H. Wapstra, Nucl. Phys. {\bf A595},  409  (1995).

\bibitem{hirsch93}
M. Hirsch, A. Staudt, K. Muto, and H.~V. Klapdor-Kleingrothaus, At. Data Nucl.
  Data Tables {\bf 53},  165  (1993).

\end{thebibliography}
\end{document}